# Effect of different buffer layers on the quality of InGaN layers grown on Si

V. J. Gómez[1,a)], J. Grandal[2], A. Núñez-Cascajero[3], F. B. Naranjo[3], M. Varela[4], M. A. Sánchez-García[2] and E. Calleja[2]

[1] *School of Physics and Astronomy, Queen's Building, 5 The Parade, Cardiff University, CF24 3AA, Cardiff, United Kingdom*

[2] *ETSIT-ISOM, Universidad Politécnica de Madrid, Avda. Complutense 30, 28040 Madrid, Spain*

[3] *Grupo de Ingeniería Fotónica, Universidad de Alcalá, Departamento de Electrónica, Alcalá de Henares, Madrid, Spain*

[4] *Grupo de Física de Materiales Complejos, Departamento de Física de Materiales & Instituto Pluridisciplinar, Universidad Complutense de Madrid. 28040-Madrid*

This work studies the effect of four different types of buffer layers on the structural and optical properties of InGaN layers grown on Si(111) substrates and their correlation with electrical characteristics. The vertical electrical conduction of n-InGaN/buffer-layer/p-Si heterostructures, with In composition near 46%, which theoretically produces an alignment of the bands, is analyzed. Droplet elimination by radical-beam irradiation was successfully applied to grow high quality InGaN films on Si substrates for the first time. Among several buffer choices, an AlN buffer layer with a thickness above 24 nm improves the structural and optical quality of the InGaN epilayer while keeping a top to bottom ohmic behavior. These results will allow fabricating double-junction InGaN/Si solar cells without the need of tunnel junctions between the two sub-cells, therefore simplifying the device design.

___________________________________

a) Author to whom correspondence should be addressed. Electronic mail: vjgomez0391@gmail.com.

**I. INTRODUCTION**

Due to the inadequacy of a single solar cell to absorb light over the full solar spectrum, a stack of multiple sub-cells (multijunction) have been proposed and studied intensively during the last few decades [1]. Most of these efforts have focused on material systems such as Ge, InP, GaAs and GaSb. However, not so much work has been reported on the use of III-Nitrides alloys to produce this type of multijunction solar cells. The wide range of band gap energies available from these alloys, their direct bandgap over the whole In content, and a high absorption coefficient, make them very attractive candidates for the multijunction solar cell approach.

Multijunction solar cells can reach energy conversion efficiencies above 30%, for the double junction case [2], or 40% for the triple one [3]. For the double-junction case, a theoretically calculated maximum efficiency of 39% can be achieved with two sub-cells with a bandgap combination of 1.74 and 1.13 eV respectively; values that can be obtained by using an InGaN-based sub-cell together with a Si one. Then, the integration of InGaN alloys with the mature Si photovoltaic (PV) technology would yield high efficiency solar cells at a reasonable cost.

One of the main problems to address in tandem cells is the current match and ohmic contact between sub-cells, yet being transparent to light absorption. Krishnamoorthy et al. demonstrated low resistance tunnel junctions (TJs) using GaN/ $In_{0.25}Ga_{0.75}N$/GaN heterostructures [4] and by introducing GdN nanoislands in an $n^{++}$-GaN/$p^{++}$-GaN junction [5]. However, TJs are quite challenging for high In content InGaN alloys because of the poor p-doping efficiency with Mg.

An additional benefit of InGaN/Si heterojunctions, for In composition around 46%, is the alignment of the n-type InGaN conduction band with the valence band of p-type silicon, that provides a low resistance ohmic contact [6]. This fact would avoid the need of TJs between sub-cells strongly simplifying the solar cell design and fabrication.

The growth of high quality InGaN alloys remains challenging due to the relatively low growth temperature of the alloy, the large lattice mismatch between InN and GaN, and hence the tendency to low miscibility and phase separation [7], [8] and [9]. Additionally, growing InGaN alloys directly on silicon, as compared to GaN/Sapphire templates, must deal with the silicon surface nitridation and the poor wetting on the resulting $Si_xN_y$ layer [10]. Silicon nitridation can be avoided by using buffer layers that may, in addition, upgrade the film quality, but should not hinder the vertical conduction in case of PV applications. This work studies the effects of different buffer layers on the vertical electrical conduction between InGaN and the silicon substrate (through the buffer layer) and correlate it with the structural and optical quality of the InGaN layer. The goal is to determine an optimal buffer layer that maximizes the InGaN crystal quality while keeping an ohmic behavior of the n-InGaN / buffer-layer / p-Si heterojunction. The quality of the InGaN layer is assessed by means of several characterization techniques such as: scanning transmission electron microscopy (STEM), electron energy-loss spectroscopy (EELS), scanning electron microscopy (SEM), atomic force microscopy (AFM), x-Ray diffraction (XRD) and cathodoluminescence at room temperature (CL). The vertical electrical conduction was derived from I-V characteristics.

## II. METHODS

All studied samples were grown on p-type Si(111) substrates in a molecular beam epitaxy (MBE) reactor equipped with a radio frequency nitrogen plasma source and standard Knudsen effusion cells for metals (Al, Ga and In).

Prior to the transfer into the growth chamber, the Si substrates were outgassed in the buffer chamber at 500 ºC during 30 min to remove water vapor and any traces of solvents. Once in the growth chamber, they were heated at 900 ºC during 30 minutes in order to remove the native oxide. The temperature was then slowly decreased (5 ºC/min) until a clear 7x7 surface reconstruction was observed (860 ºC) at the RHEED pattern [13]. This procedure was followed regardless of the buffer layer subsequently grown in the MBE system.

The buffer layers employed in this study were the following:

AlN layers grown by MBE at 860 ºC with nominal thickness of 6, 24, 42 and 84 nm. A 6 nm thick $In_{0.10}Al_{0.90}N$ layer grown by MBE at 550 ºC. RF-sputtered $In_{0.39}Al_{0.61}N$ layers grown at room temperature (RT) with nominal thickness of 5 and 20 nm. Intentionally nitridated Si at 860 ºC to generate a 2-3 nm thick $\beta$-$Si_xN_y$ layer.

The AlN buffers were grown first depositing 4.5 nm of Al on the Si(111) surface at 860 ºC and then turning on the plasma source, so that the spontaneous formation of $Si_xN_y$ was prevented [14], [15]. The AlN buffers were then grown at the same temperature, with Al and N fluxes of $7.0x10^{14}$ $at/cm^2s$ and $7.2x10^{14}$ $at/cm^2s$ respectively. A streaky 1x1 RHEED pattern was observed at the end of the AlN buffer growth, indicating a smooth and flat surface.

A nominal 6 nm thick $In_{0.10}Al_{0.90}N$ buffer layer was grown by MBE at 550 ºC to study the influence of adding In to an AlN buffer layer, which should improve the electrical conductivity. The buffer growth also started with a 4-5 nm of Al pre-deposition at 550 ºC. The fluxes were set to $\phi_{Al} = 7.2x10^{14}$ $at/cm^2s$, $\phi_{In} = 3.2x10^{14}$ $at/cm^2s$ and $\phi_{N*} = 1.2x10^{15}$ $at/cm^2s$. A spotty RHEED pattern, typical of 3D-growth, was observed during the whole growth. The In composition was set to 10% due to the inherent difficulty to incorporate In in the alloy at such high temperature (550 ºC) [16]. Lower temperatures would not be adequate for Al containing alloys.

Polycrystalline $In_{0.39}Al_{0.61}N$ layers 5 and 20 nm thick were deposited by rf-sputtering at RT on Si (111) substrates. Rf-sputtering was employed due to its capability to deposit high In content InAlN alloys with a reasonable crystal quality. The In composition was selected to reduce the lattice mismatch between the InGaN epilayer and the buffer, from 5% (AlN) to only 1%. According to [18] sputtered $In_{0.36}Al_{0.64}N$ showed RT-PL emission at 1.75 eV. A further increase on the In content of the InAlN buffer layer to reduce the lattice mismatch between the InGaN layer and the InAlN buffer would result in a reduction on the photons that arrive at the hypothetic Si bottom cell ($E_G$ = 1.11 eV at RT); thus, reducing the overall efficiency of the device.

Intentional nitridation of the Si substrate (with no Al predeposition) at 860 ºC during 5 min under $\phi_{N*} = 1.2x10^{15}\ at/_{cm^2 s}$ generated a 2-3 nm thick β-$Si_xN_y$ layer. According to [17] a continuous, reproducible and somehow crystalline β-$Si_xN_y$ layer is formed when nitridation is performed at high enough temperatures (850 ºC). The crystal quality of the III-Nitride epilayer grown on it drops dramatically when the $Si_xN_y$ is produced at either intermediate (450 ºC) or low (150 ºC) temperature [17]. In these cases the $Si_xN_y$ layers seems to be amorphous. Under proper nitridation conditions (above the 7x7 to 1x1 transition temperature), the Si surface is converted into crystalline β-$Si_xN_y$ (0001) [19], [20] that may be lattice matched to Si(111) [19]. Rawdanowicz et al. suggested that amorphization of the β-$Si_3N_4$ layer occurs during the TEM specimen preparation, due to the $Ar^+$ ion milling [21], the reason why these layers, when inspected by TEM, appear as amorphous.

For the growth of the InGaN layers, the Ga flux was varied between $4.5x10^{14}$ and $6.8x10^{14}\ at/_{cm^2 s}$; and the In flux from $3.2x10^{14}$ to $7.4x10^{14}\ at/_{cm^2 s}$. The nitrogen fluxes employed were $6.7x10^{14}$ and $1.2x10^{15}\ at/_{cm^2 s}$. All the InGaN layers described in this study were grown using the advanced-DERI method [11] and [12] at a substrate temperature of 550 ºC. Advanced-DERI consists of two series of growth processes: a metal-rich growth process (MRGP) and a droplet elimination one (DEP). During the MRGP, InGaN is grown under metal-rich conditions ($\phi_{Ga} + \phi_{In} > \phi_N$) with the condition that $\phi_{Ga} < \phi_N$. Ga is preferentially captured and excess In is easily swept out to the surface forming droplets. These droplets are eliminated by epitaxial transformation to InGaN during the DEP by shutting off the In flux and keeping the prior condition of $\phi_{Ga} < \phi_N$. Once all the droplets are eliminated (observed by RHEED intensity) the MRGP starts again. It is important to maintain always an excess of In on the surface to avoid reduction on the In content and pure GaN inclusions. The optimal conditions to grow the InGaN films in this work were 2 minutes of MRGP and 1 minute of DEP, with a repetition of either 19 or 30 times (depending on the nitrogen flux used) to keep the same InGaN thickness (~0.5 µm). Up to the best of our knowledge, all previous reports on InGaN layers grown by DERI were done on GaN/Sapphire or on InN/GaN/Sapphire; no reports are found on Silicon substrates.

### III. RESULTS AND DISCUSSION

#### A. Surface structure, composition and morphology

Two of the most influential characteristics of the buffer layers are the coverage and the crystal orientation. Buffer layers covering partially the substrate will let areas exposing bare silicon were the InGaN epilayer may also grow with different quality, blurring somehow the overall effect of the buffer.

For this purpose, the structural and chemical composition of a 42 nm thick AlN buffer layer was analyzed by plan-view TEM and EELS respectively. STEM-EELS measurements were carried out in an aberration-corrected JEOL ARM200cF electron microscope operated at 200 kV and equipped with a cold field emission gun and a Gatan Quantum EEL spectrometer. Samples were prepared by conventional mechanical grinding and Ar ion milling. In addition, the morphology of both InGaN films and the buffer layers were inspected by SEM and AFM. The differences in InGaN surface morphology are significant and very dependent on the buffer layer employed for the range of In contents employed in this work.

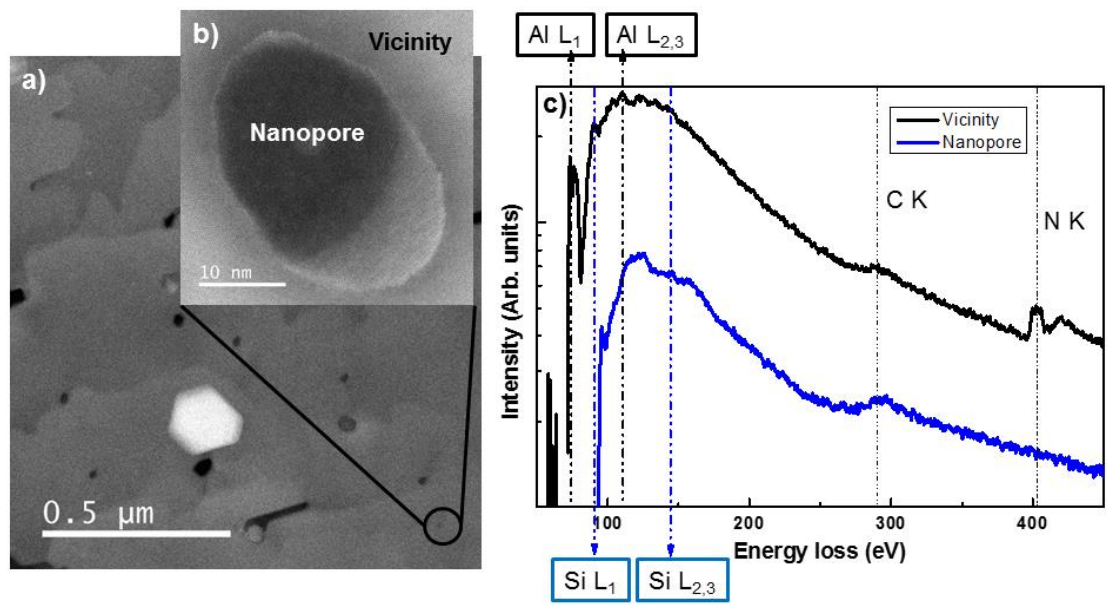

**FIG. 1.** a) ADF micrograph of the 42 nm thick AlN buffer layer. b) ADF micrograph of a nanopore. c) EEL spectra inside the nanopore (blue) and on its vicinity (black).

In figure 1a the plan-view high angle annular dark field (ADF) micrograph of the 42 nm-thick AlN buffer shows sub-micron size grains, together with a significant density of nanopores with diameters ranging from 10 to 20 nm (fig. 1b shows one nanopore magnificated). EELS measurements identified the elements present inside and outside (in the vicinity) the nanopore. Inside the nanopore (fig. 1c, bottom curve) the silicon L1, L2 and L3 edges are the only ones observed indicating that the nanopore threads down to bare silicon. The nitridation of the Si surface at the bottom of the nanopores seems to be avoided by the shadowing effect produced by the close nanopore walls. The EEL spectrum taken at the nanopore vicinity (fig. 1c, top curve) shows the lines corresponding to the aluminium L1, L2, L3 edges and the nitrogen K edge, convoluted with those from silicon, indicating that AlN is grown on Si. The 284 eV edge that appears on both spectra corresponds to the carbon K edge.

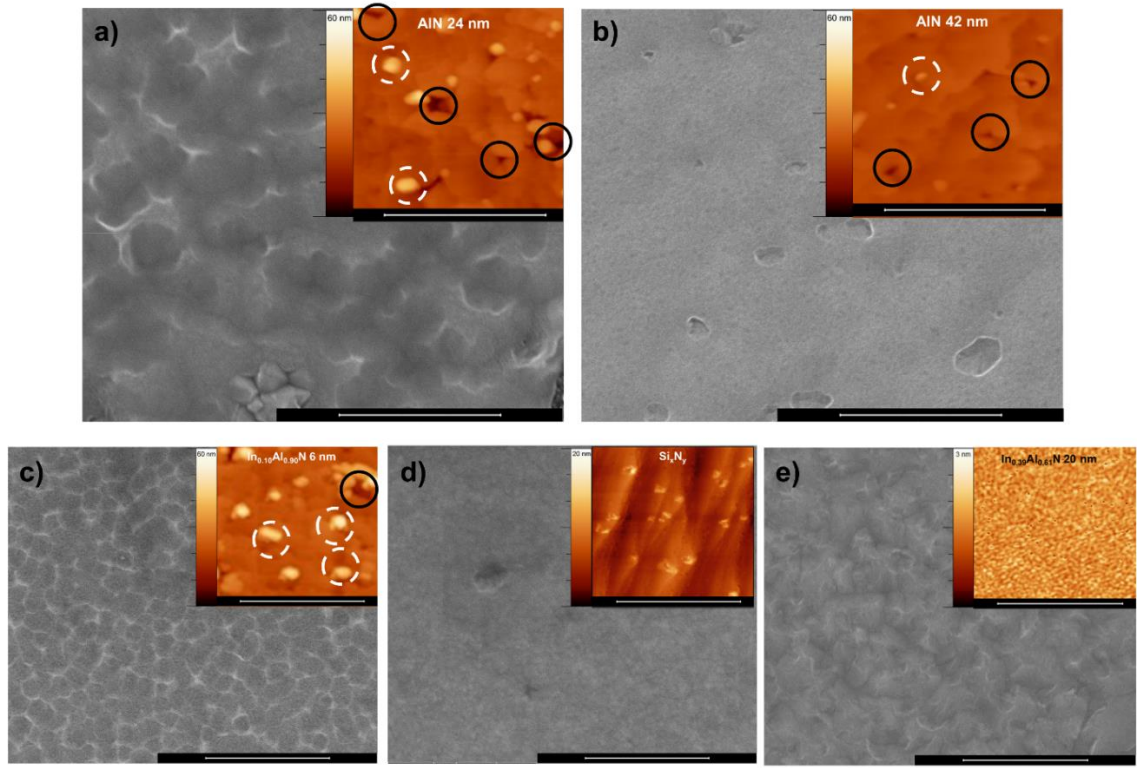

**FIG. 2.** Top view SEM images of InGaN layers grown on: a) 24 nm AlN buffer, b) 42 nm AlN buffer, c) 6 nm In$_{0.10}$Al$_{0.90}$N MBE, d) Si$_x$N$_y$ and e) 20 nm sputtered In$_{0.39}$Al$_{0.61}$N. Inset shows AFM images of the starting buffer layer. The black circles mark some of the nanopores, and the dashed circles the macroscopic defects. In all cases, the scale bar is 1 µm.

SEM micrographs of the InGaN epilayers surfaces grown on top of the different buffers are shown in figure 2. Depending on the buffer layer used, different features are observed. In the case of the AlN and InAlN MBE buffers (inset of the figures 2a, b and c) the surface is covered by macroscopic defects (white dashed circles in the inset of figures 2a, b and c), and nanopores (black circles in the inset of figures 2a, b and c). The density of the nanopores is higher for the case of the 24 nm AlN than for the 42 nm AlN and 6 nm InAlN (MBE), and the diameter of them is lower for the 42 nm AlN buffer. The density of surface defects is higher for the case of the 6 nm InAlN (MBE) than for the 24 nm AlN and 42 nm AlN. Due to the presence of nanopores in the buffer layer, the InGaN islands can nucleate directly on the (In)AlN or on the bare Si at the bottom of the nanopores. The islands continue growing, both vertically and laterally, until they coalesce into a compact and grainy layer. The effect of a lower density of nanopores is a reduction on the parasitic growth of InGaN on bare Si, yielding smoother InGaN epilayers, as in the case of the InGaN layers grown on a 42nm AlN.

The intentional nitridation of the Si surface under optimum conditions (temperature over 860 ºC [17]) produces a thin (2-3 nm) β-Si$_x$N$_y$ layer [19], [20]. The InGaN epilayer grown on top shows a smooth surface (fig. 2d), comparable to the one growth on the 42 nm AlN (fig. 2b). The polycrystalline sputtered InAlN buffer presents a grainy surface which leads to rough InGaN epilayers (fig. 2e).

**B. Compositional uniformity**

The compositional uniformity (at macroscopic level) of the InGaN films was assessed by symmetric ω/2θ scans around the Si (111) Bragg reflection. Figure 6.3 shows the ω/2θ diffraction profiles of all samples, where the Si (111) and InGaN (0002) reflections are shown in all cases. The reflections from the buffer layers could only be measured for AlN with thickness above 24 nm. The 6 nm AlN and InAlN epitaxial buffers are too thin to distinguish any diffraction peak from the noise level. Neither the $Si_xN_y$ interlayer nor the polycrystalline sputtered InAlN show any diffraction peak.

All InGaN layers, independently of the buffer used show similar InGaN (0002) single diffraction peaks. The absence of double diffraction peaks together with the relatively narrow InGaN reflections confirm the absence of significative (measurable) phase separation. This result is quite interesting because avoiding phase separation in InGaN alloys is challenging, especially when growing alloys close to the low miscibility gap (30% to 70% of In) and with a large mismatch with the substrate (>8%). All InGaN layers are partially relaxed (>80%) as derived from RSM data around the Si(224) Bragg reflection (data not shown here).

All InGaN diffraction peaks deviate from an ideal Lorentzian symmetry showing a tail extended towards lower indium contents, which may be attributed to the compositional pulling effect at the early stages of the growth. This effect, which has been observed in both nanowires [23] and thin films [24], [25], happens to ease the lattice mismatch between the InGaN layer and the buffer by reducing the In incorporation.

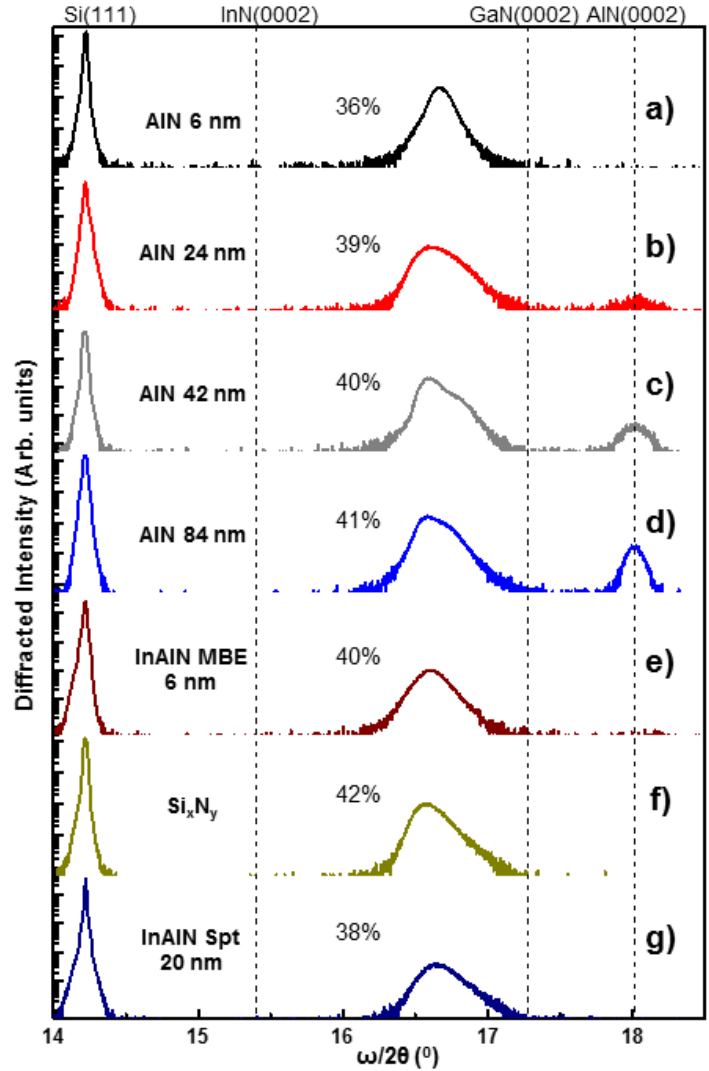

**FIG. 3.** Symmetric ω/2θ diffraction profiles around the Si(111) Bragg reflection are shown for the InGaN layers grown on a) 6 nm, b) 24 nm, c) 42 nm and d) 84 nm thick AlN, e) 6 nm thick InAlN, f) $Si_xN_y$ and g) 20 nm sputtered InAlN. In all cases InGaN(0002) Bragg reflection is shown.

## C. Crystallographic uniformity

ω-rocking scans were recorded around the InGaN (0002) reflection in order to determine the InGaN layers crystallographic uniformity (in terms of the dispersion on the alignment of the c-planes of the InGaN epilayer). Figure 4 shows the ω-rocking (0002) full width at half-maximum (FWHM) value versus the In content for all InGaN layers. The samples could be divided into three groups depending on the buffer layer used: i.- (In)AlN MBE buffer layers, ii.- $Si_xN_y$ buffer and iii.- InAlN sputtered buffer layers. InGaN layers grown on MBE (In)AlN epitaxial buffers have the lowest FWHM values, followed by those grown on $Si_xN_y$ layers (~2.5 times higher), to end up with those grown on sputtered InAlN which is the worst case (~5.5 times higher). This result is not very surprising since low FWHM values are expected from the highest buffer crystal quality that also leads to an epitaxial relation with the epilayer. On the other hand, the polycrystalline nature of the sputtered InAlN buffer induces a similar misorientation on the grown InGaN layer that also becomes polycrystalline. The InGaN epilayers grown on $Si_xN_y$ have a FWHM of ~2.5º. This value is lower than the ones obtained for the layers grown on sputtered InAlN. This is most likely due to the high nitridation temperature (860 ºC), that is known to form a somehow crystalline or polycrystalline (β-)$Si_xN_y$ layer [17]. As can be seen in figure 4, the type of buffer has a deeper impact on the FWHM value than the indium content (within the range studied), or the growth conditions.

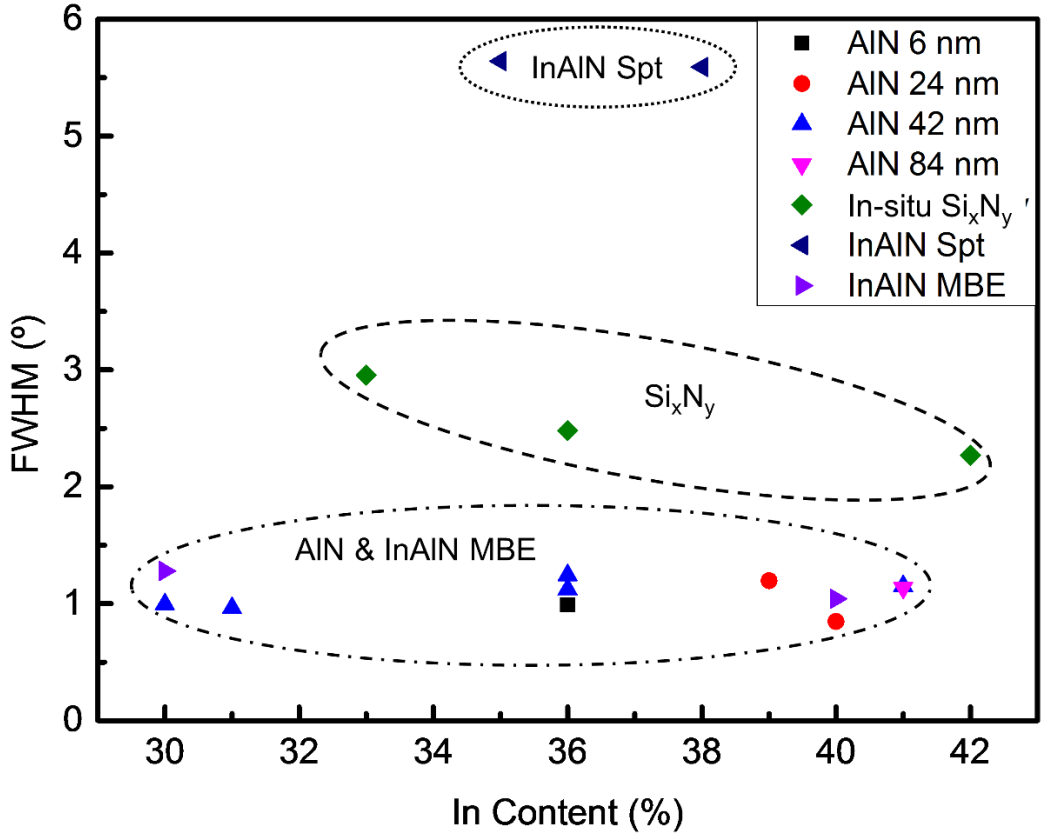

**FIG. 4.** ω-rocking FWHM of the InGaN (0002) Bragg reflection vs In content for the different buffer layers used.

**D. Epitaxial relationship**

The epitaxial relationship between the silicon substrate and the InGaN epilayer was investigated by phi-psi pole figures around the Si (220) Bragg reflection. This configuration was selected due to the similar ω/2θ values for the Si(220) (ω/2θ = 23.65º/47.30º) and for the InGaN(10-12) (ω/2θ ~ 23.3º/46.6º). The pole maps were recorded in the ranges Ψ = 30.00 - 50.00º and ϕ = 0 - 360º. The four pole maps of figure 5 show the three-fold symmetry of the Si(220) planes. The six-fold symmetry (typical of wurtzite structures) of the InGaN(10-12) is shown for the InGaN layers grown on AlN (fig. 5a), epitaxial InAlN (fig. 5b) and SixNy (fig. 5c). There is no evidence of InGaN related peaks when the sputtered InAlN buffer is used. The peaks corresponding to epitaxial (In)AlN buffers are not observed in any pole map. This is most likely due to its low thickness together with the fact that the position of the (In)AlN (10-12) reflection is coincident with the InGaN (10-12) one. The epitaxial relationships for InGaN layers grown on epitaxial (In)AlN buffers and on SixNy are determined by the ω/2θ scans and pole maps as: InGaN (0001)∥Si (111) and InGaN (10-10)∥Si (11-2).

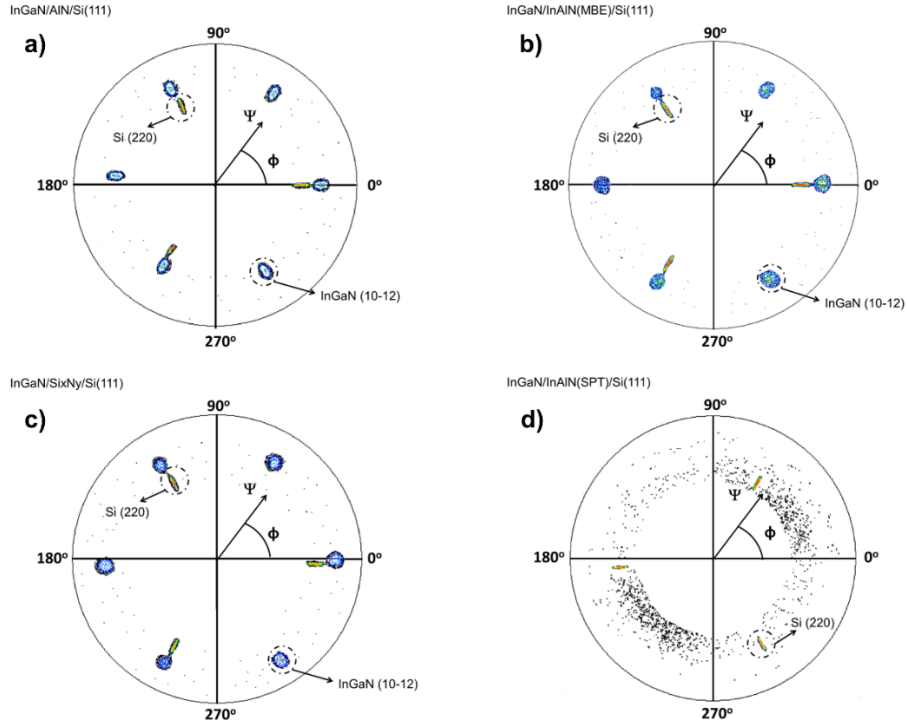

**FIG. 5.** φ-Ψ pole maps around the Si(220) Bragg reflection for the InGaN layers grown on a) AlN, b) epitaxial InAlN (MBE), c) $Si_xN_y$ and d) Sputtered InAlN.

The intentional nitridation of the Si surface, at high enough temperatures, produces a somehow crystalline or polycrystalline β-$Si_xN_y$ layer. Epitaxial relationship between InGaN and Si(111) when using $Si_xN_y$ interlayer has been widely observed when growing GaN [17], [26], and InGaN [10] on $Si_xN_y$ / Si. In this work, at the conditions selected for the nitridation the Si surface is converted into crystalline β-$Si_xN_y$(0001) [19], [20] that may be lattice matched to Si(111) [19]. This is, most likely, the reason why there is an epitaxial relationship between the InGaN layer and the silicon substrate. Another possible explanation was proposed by Tamura et al. [27]. They proposed that the thin $Si_xN_y$ layer covered partially the Si surface, thus leaving some holes with bare Si that allowed an epitaxial relation between the III-Nitride film and the Si substrate. The III-Nitride islands nucleated on those holes would eventually overcome the SiN thickness and start growing laterally (similar to an ELOG process) thus providing a rather high quality film. However, up to the best of our knowledge, no experimental evidence of the existence of these "pin-holes" has been reported, when nitridating the silicon surface inside the MBE growth chamber. The determination of this mechanism is not the scope of this work.

The pole figure from the InGaN layer grown on sputtered InAlN (fig. 5d) reveals the three peaks corresponding to the Si(220) planes, while the six peaks related to InGaN are not detected. As mentioned before, it is assumed that the polycrystalline nature of this buffer promotes the growth of polycrystalline InGaN. Due to the random in-plane orientation of the grains (twist) in polycrystalline layers, signals neither from InGaN nor from sputtered InAlN can be seen in the pole map. Despite of being polycrystalline, the InGaN layer is mostly c-plane oriented as determined from ω/2θ scans (fig. 3g).

**E. Cathodoluminescence**

The analysis of the optical properties of InGaN layers was carried out by means of RT-CL (figure 6). Measurements at RT are very sensitive to the presence of non-radiative recombination centers. Out of the series of grown InGaN layers on (In)AlN buffers, the one grown on the thinnest AlN buffer (6 nm) does not show CL emission (fig. 6a), most likely due to a high density of non-radiative recombination centers. The other InGaN layers, grown on thicker AlN buffers, show CL emission with comparable intensity (less than a factor of two) pointing to a rather good optical quality. The lowest optical quality is obtained in InGaN layers grown on epitaxial InAlN buffers (fig. 6e).

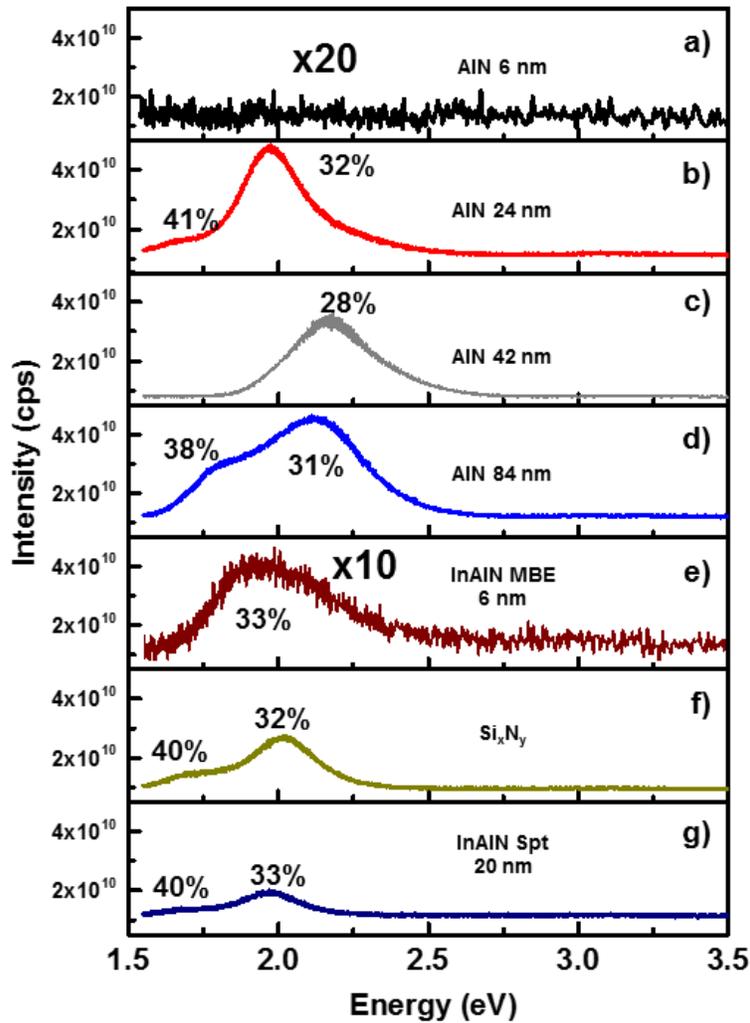

**FIG. 6.** Cathodoluminescence spectra (obtained at RT) of the InGaN layers grown on: a) 6 nm, b) 24 nm, c) 42 nm, and d) 84 nm thick AlN, e) 6 nm thick InAlN, f) $Si_xN_y$ and g) sputtered InAlN.

InGaN layers grown on SixNy (fig. 6f) and on sputtered InAlN (fig. 6g) show CL emission intensity comparable to the layers grown on AlN buffers (above 6 nm thick) (figs. 6b, c and d). The shoulders and double peaks that appear in the PL spectra point to an inhomogeneous In composition that may be attributed to the mentioned compositional pulling effect.

Dobrovolskas et al., [28] studied by PL InGaN layers grown by DERI finding two PL bands originating at regions with different InGaN compositions: one near the substrate with lower In% and another near the surface with a higher In%. These results are in good agreement with lattice pulling effect and indicate that the lower energy emission in figure 6 corresponds to the upper region in the InGaN layer. When only one of the emissions is present, it could be due to a higher density of defects of that InGaN sub-layer. The In% values given in figure 6 were estimated using a bowing parameter of 2.5 eV [22] and taking into account the strain of the InGaN layer (measured by RSM).

**F. Electrical characterization**

The samples were electrically characterized by measuring the I-V characteristic between vertical contacts (fig. 7). Ti/Al/Ni/Au contacts were deposited on top of the non-intentionally doped (NID) n-type InGaN layers (residual electron density of ~$10^{18}$ cm$^{-3}$) grown on different buffers and annealed at 400 ºC during 5 minutes to form an ohmic contacts, while Al/Au was deposited on the backside of the p-type Si substrate. Horizontal pairs of contacts were used on the InGaN layers and p-type substrates to verify their ohmic behavior and to calculate the contact resistivity (~10 $\Omega$cm$^2$ in case of contacts on p-Si and ~20 $\Omega$cm$^2$ for contacts on InGaN).

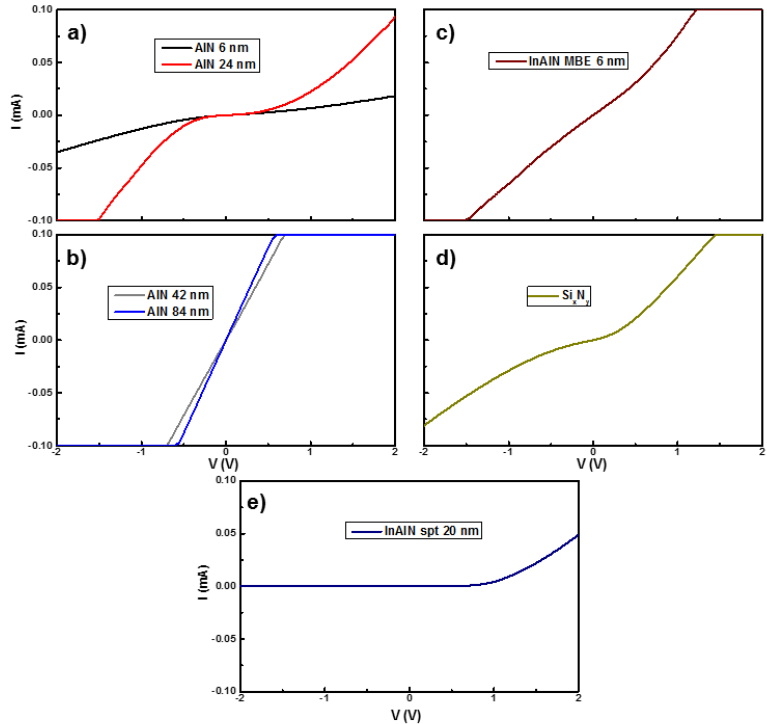

**FIG. 7.** Vertical I-V curves of samples grown on a) 6 and 24 nm, b) 42 and 84 nm thick AlN, c) 6 nm thick InAlN, d) Si$_x$N$_y$ and e) 20 nm thick sputtered InAlN.

Figures 7a and 7b show the I-V characteristics of the InGaN layers grown on the thin (6 and 24 nm) and thick (42 and 84 nm) AlN buffer layers. In the case of the thin AlN, a non-ideal rectifying behavior is observed, whereas for the thick ones it is ohmic. From the results obtained in previous sections, thin AlN buffer layers present a large density of nanopores, which can induce the growth of InGaN directly on the bare Si substrate together with the growth on the corresponding buffer layer. In any case, one may expect a highly defective interface, which would prevent the ideal electrical conduction. However, for the case of thick (42-84 nm) AlN buffer layers (Figure 7b) an ohmic behavior is observed. This type of buffer layer presented a lower density of nanopores and smaller diameter, leading to a larger size of the grains.

Two current flow mechanisms are proposed: one through the nanopores and another through grain boundaries. The former is the dominant for the thin AlN (6 and 24 nm) (fig. 7a), in which the density of nanopores is one order of magnitude higher and its diameter is between 2 and 5 times larger as compare to the thick (42 and 84 nm) AlN buffer layers. In addition, the poorer coverage of the Si surface by the AlN buffer layer, leads to less connection between grains and thus to less grain boundaries. Both facts together make the conduction through nanopores the dominant mechanism for the mentioned structures. For the case of thick AlN buffer layers the dominant current flow mechanism is related to the grain boundaries; the nanopores density and diameter are both decreased by 10 and 2-5 times, respectively, together with the larger size of the grains that allow an increase in the number of grain boundaries. Grain boundaries are known to facilitate charge transfer across insulating layers [29]. This strong transport property would not be achieved if the AlN buffer were both compact (absence of nanopores) and monocrystalline. In that case, the charge transport could only take place by tunneling, which would not be allowed at the measured currents. To support this explanation, Musolino et. al. [30] reported electroluminescence of GaN nanowires grown on AlN buffered Si. They concluded that the grain boundaries present in the AlN buffer allows the current to flow through them.

The results obtained for the other three buffers cannot be explained in the same way due to the intrinsic differences between buffers. Further investigations should be performed in order to clarify those results.

## IV. CONCLUSIONS

In summary, we have identified that the use of an AlN buffer layer (of thickness between 42 and 84 nm) leads to an improvement of the structural and optical properties of the InGaN epilayer, while keeping the ohmic behavior of the heterointerface. This could help in a possible integration of III-nitrides with Si solar cells by avoiding the use of a tunnel junction, simplifying the design and fabrication of the device. Several transport mechanisms have been suggested to understand the differences in electrical behavior. However, focusing on AlN buffers, transport through nanopores and grain boundaries seem to be most likely the ones plating a role, though there is no direct evidence. In addition, the successful growth of InGaN films by DERI on buffered Si substrates is achieved and reported for the first time.

**ACKNOWLEDGMENTS**

The authors acknowledge Ms. M.C. Sabido for her work in processing the electrical contacts, and Dr. P. Aseev for fruitful discussions. The work was partly supported by the Spanish Ministry of Science and Innovation, project MAT2011-26703 and ETSIT UPM. Research at UCM sponsored by MINECO/Feder grant MAT2015-066888-C3-3-R and European Research Council PoC2015 grant "MAGTOOLS", GA# 713251.